\documentclass[aps,showpacs,preprint,superscriptaddress,12pt]{revtex4}
\usepackage{graphicx}
\usepackage{bm}
\usepackage{txfonts}
\usepackage{subfigure}

\begin{document}
\title{Scale invariance and scaling law of Thomson backscatter spectra by electron moving in laser-magnetic resonance regime}
\author{Yi-Jia Fu}
\affiliation{College of Nuclear Science and Technology, Beijing Normal University, Beijing 100875, China}
\author{Chong Lv}
\affiliation{College of Nuclear Science and Technology, Beijing Normal University, Beijing 100875, China}
\author{Feng Wan}
\affiliation{College of Nuclear Science and Technology, Beijing Normal University, Beijing 100875, China}
\author{Hai-Bo Sang}
\affiliation{College of Nuclear Science and Technology, Beijing Normal University, Beijing 100875,  China}
\author{ Bai-Song Xie\footnote{Corresponding author. Email address: bsxie@bnu.edu.cn}}
\affiliation{College of Nuclear Science and Technology, Beijing Normal University, Beijing 100875,  China}
\affiliation{Beijing Radiation Center, Beijing 100875, China}

\date{\today}
\begin{abstract}
The Thomson scattering spectra by an electron moving in the laser-magnetic resonance acceleration regime are computed numerically and analytically. The dependence of fundamental frequency on the laser intensity and magnetic resonance parameter is examined carefully. By calculating the emission of a single electron in a circularly polarized plane-wave laser field and constant external magnetic field, the scale invariance of the radiation spectra is evident in terms of harmonic orders. The scaling law of backscattered spectra are exhibited remarkably for the laser intensity as well for the initial axial momentum of the electron when the cyclotron frequency of the electron approaches the laser frequency. The results indicate that the magnetic resonance parameter plays an important role on the strength of emission. And the rich features of scattering spectra found may be applicable to the radiation source tunability.

\end{abstract}
\maketitle

\section{INTRODUCTION}

The significant and continuous development of the chirped-pulse amplification technique \cite{SCi-264-917} has intrigued many researches in the high field science with rich nonlinear features \cite{Nat-333-337,PRL-85-570,PRE-52-5443,PRL-88-055004,Natp-2-696}. Particularly the laser-matter interaction has also provided the possibility to achieve the advanced x-ray sources. Various schemes have been proposed and developed for the compact and tunable x-ray sources, for example, the high harmonic generation (HHG) \cite{Nat-406-164,POP-8-1774}, the Thomson and Compton scattering \cite{Nat-333-337,PRL-85-570,PRE-52-5443,PRL-88-055004,Natp-2-696,APL-103-174103} and the free electron laser (FEL) \cite{Natp-4-641} and so on.

The generation of x rays by the Thomson scattering, i.e., the scattering spectrum produced by the relativistic electron in the high-intensity laser field, exhibits the rich structure and even irregular distribution characteristic.
Several studies \cite{PRA-54-4383,JPA-30-4399,PRA-55-3678,PRA-58-3221,PRL-80-5552} about the Thomson-scattering are focused on the characteristics of spectrum dependence of an external uniform magnetic field. The external magnetic field not only confines the transverse motion of the electron but also distorts the laser field so that the resonance absorption of emission \cite{PRL-80-5552} occurs.

Most of former researches have been based on the assumption that the frequency of harmonic radiation is an integer multiple of the laser frequency, however, this assumption is not valid when the laser field as well an external magnetic field are both presented. Actually the perspective of the laser-magnetic resonance acceleration \cite{POP-12-053103,JAP-100-044907,APL-87-254102,PRE-68-046407,PRE-69-066409,PRE-66-036406,APL-95-161105} tells us that
the relativistic cyclotron motion will occur in transverse direction under an external magnetic field. This cyclotron-resonance motion will affect the frequency of emission. For example the authors in Ref.\cite{PRE-66-036406} demonstrates that the electron will emit radiation at high harmonics of the cyclotron frequency in the plane of the electron cyclotron orbit. Beside, when the cyclotron-resonance condition is met, i.e., the cyclotron frequency approaches to the laser frequency, the periodic motion electron makes would result in that the fundamental frequency of harmonic emission is neither the laser frequency nor the cyclotron frequency for the backscattered Thomson spectra \cite{POP-9-4325}. Evenly the impact of cyclotron-resonance degree of the electron motion upon the radiation spectrum can be adjusted by changing the electron orbit.
Since the phenomena of Thomson scattering spectra of moving electron in the combined laser-magnetic fields are very rich and highly nonlinear and some influences of factors are not included in previous researches, on the other hand, motivated by the advantage of electron resonance acceleration in laser field assisted by the external magnetic field in our previous study \cite{APL-95-161105}, we study the involved Thomson scattering in presence of laser and magnetic field.

In this paper, we consider the motion of the electron in circularly polarized laser-plane wave and static magnetic field. Different from the former works \cite{PRA-60-3276,PRA-60-2505}, our analytic solution of the motion equation indicates that the electron makes the strict periodic motion, therefore, the fundamental frequency of harmonic radiation would be affected by the combination of factors which include the period of the electron motion, the direction of observation and the electron orbit rather than only the laser frequency or the cyclotron-resonance frequency studied in Ref.\cite{POP-10-2155,POP-9-4325}. On the other hand,it is found that the fundamental frequency of harmonic emission is independent of the initial phase of the electron with respect to the laser field, this point is contrast to the work by He \cite{POP-9-4325}. By the way the magnetic resonance parameter defined in magnetic resonance acceleration, see our previous work \cite{APL-95-161105}, which plays an important role in our studied problem, is used to present the relationship between the degree of cyclotron resonance and the emission spectra. The main finding of the present work is the scale invariance of the backward scattering spectrum on the basis of integer harmonic orders especially the laser intensity and the magnetic resonance parameter are used as scale parameters in highly cyclotron resonance cases. By the theoretical analysis and numerical results, the scaling laws of magnetic resonance parameter as well the initial axial momentum are found and its implications are discussed.

The paper is organized as follows. In Sec. \ref{basic}, the momentum and trajectory of the electron in the circularly laser-plane field and constant external magnetic field are derived from the parametric equations of the normalized laser intensity, the magnetic resonance parameter and the phase of the electron. The energy of the electron is also given in the parametric form. The feature of the radiation spectrum via the analytic formula is given. In Sec. \ref{numeric}, the backscattered spectrum is obtained numerically and compared with analytic results in previous section. Our main conclusions and a brief discussion will be given in the final section.

\section{basic equations}\label{basic}

In the present paper, we consider the problem of Thomson scattering by an electron (with mass $m$ and charge $-e$) moving in an intense plane wave and an external uniform magnetic field. It is assumed that the plane wave, with vector potential amplitude $A_{0}$ and frequency $\omega_{0}$, is circularly polarized and propagates in the positive $z$ direction. An external uniform magnetic field $B_{0}$ is also along the positive $z$ direction. By denoting the phase of the laser field as $\eta={\omega}_{0}t-\textit{\textbf{k}}\cdot\textit{\textbf{r}}$, where $\textit{\textbf{k}}$ and $\textit{\textbf{r}}$ is the laser wave vector and electron displacement vector, respectively, then the combinational total vector potential of laser and uniform magnetic fields can be expressed as
\begin{equation}
\label{eq1}\textit{\textbf{A}}=
{A}_{0}\left [-\sin {\eta}\hat{\textbf{i}}+\cos{\eta}\hat{\textbf{j}}\right ]+{B}_{0}x\hat{\textbf{j}},
\end{equation}
therefor, the corresponding electric field $\textit{\textbf{E}}$ and magnetic field $\textit{\textbf{E}}$ will be derived from the vector potential $\textit{\textbf{A}}$ via the equations
\begin{eqnarray}
\label{eq2}\textit{\textbf{E}}=-\frac{1}{c}\frac{\partial\textit{\textbf{A}}}{\partial t},\\
\textit{\textbf{B}}=\triangledown \times \textit{\textbf{A}}.
\end{eqnarray}
The dynamics of electron will be studied on the basis of the following energy-momentum transfer equations:
\begin{equation}
\label{eq3}\frac{d \textit{\textbf{p}}}{d t}=-e\left(\textit{\textbf{E}}+\boldsymbol{\beta}\times \textit{\textbf{B}}\right),
\end{equation}
and
\begin{equation}
\label{eq4}\frac{d\left( \gamma mc^{2}\right) }{d t}=-ec\boldsymbol{\beta }\cdot\textit{\textbf{E}},
\end{equation}
where $\textit{\textbf{p}}$ is the electron relativistic momentum, $\boldsymbol{\beta}$ is the electron velocity (in units of $c$) and $\gamma=\left(1-\beta^{2}\right)^{-1/2}$ is the electron relativistic factor. For convenience, we normalize time by $1/\omega_{0}$, velocity by $c$ and distance by $k_{0}^{-1}=c/\omega_{0}$. The potential and magnetic field are measured by dimensionless parameters $a=e\textit{\textbf{E}}_{0}/m\omega_{0}c$ and $b=e\textit{\textbf{B}}_{0}/m\omega_{0}c$, respectively.

\subsection{The trajectory solutions}

Since that an electron moves in constant circularly laser light, the constant of motion is readily obtained as $\varsigma =\gamma-p_{z}=\gamma_{0}-p_{z0}$. From Eqs. (\ref{eq3}) and (\ref{eq4}), we obtain the equations of electron motion
\begin{eqnarray}
\label{eq5}\frac{d^2 p_{x} }{d \eta ^2}+{\omega _{b}}^2 p_{x}= (\omega_{b} +1)a\sin{\eta},\\
\label{eq6}\frac{d^2 p_{y} }{d \eta ^2}+{\omega _{b}}^2 p_{y}=- ( \omega_{b} +1)a\cos{\eta },
\end{eqnarray}
where $\omega_{b}= b/ \varsigma$ is the modified cyclotron frequency of the electron motion in the combined magnetic and laser fields. We assumed that the phase $\eta$ has an initial value $\eta_{in}=-z_{in}$ at $t=0$. In order to describe the electron energy gain in the laser field and also the Thomson scattering spectra in the following we define the resonance parameter as $n=1/\left |\omega_{b}-1\right|$.
Obviously, more approaching to $\omega_{b}=1$ when resonance condition satisfies, more high value of $n$ is. Thus we call it a high degree resonance when $n \gg 1$.

By using Eqs.(\ref{eq5}-\ref{eq6}), we will obtain the momentum and energy of the electron as
\begin{eqnarray}
\label{eq7}p_{x}=na\left \{ \sin{\eta}-\sin{\left[\omega_{b}\eta-\left( \omega_{b}-1\right)\eta_{in}\right]} \right \},\\ \label{eq8}p_{y}=na\left \{ -\cos{\eta}+\cos{\left[\omega_{b}\eta-\left( \omega_{b}-1\right)\eta_{in}\right]} \right \},\\
\label{eq9}p_{z}=\frac{2n^2a^2}{\varsigma}\sin^2{\left[\frac{\left(\omega_{b}-1 \right )\left(\eta-\eta_{in} \right )}{2}\right]}+\frac{1}{2\varsigma}-\frac{\varsigma}{2},\\
\label{eq10}\gamma=\frac{2n^2a^2}{\varsigma}\sin^2{\left[\frac{\left(\omega_{b}-1 \right )\left(\eta-\eta_{in} \right )}{2}\right]}+\frac{1}{2\varsigma}+\frac{\varsigma}{2}.
\end{eqnarray}
From Eqs.(\ref{eq7}-\ref{eq10}), it is evident that the momentum and energy of the electron have a singularity at $b=\varsigma$, i.e., $\omega_{b}=1$ in the strict cyclotron resonance condition. By the way the energy may have two resonant points $b=\pm\varsigma$ that correspond to the left-hand and right-hand circularly polarized laser field \cite{PRE-69-066409}. For simplicity we only care about the plus situation.
Obviously from the analytical formula of the transverse momentum and the energy it is found that the transverse momentum will increase as $n$ by the order of $n$, meanwhile the energy will increase by $n^{2}$.

Finally the equations of the electron trajectory can be obtained via $d \boldsymbol{r}/d \eta=\boldsymbol{p}/\varsigma$ as follow
\begin{eqnarray}
\label{eq11}x(\eta)=\frac{ na}{\varsigma}\left \{ -\cos{\eta}+\frac{1}{\omega_{b}}\cos{\left[\omega_{b}\eta-\left ( \omega_{b}-1 \right )\eta_{in} \right ]} -\left ( \frac{1}{\omega_{b}}-1 \right )\cos{\eta_{in}}\right \},                                        \\
\label{eq12}y(\eta)=\frac{na}{\varsigma}\left \{ -\sin{\eta}+\frac{1}{\omega_{b}}\sin{\left[\omega_{b}\eta-\left ( \omega_{b}-1 \right )\eta_{in} \right ]} -\left ( \frac{1}{\omega_{b}}-1 \right )\sin{\eta_{in}}\right \},                                             \\
\label{eq13}z(\eta)=\left ( \frac{na}{\varsigma} \right )^{2}\left \{ \left ( \eta-\eta_{in} \right )-\frac{1}{\omega_{b}-1}\sin{\left [ \left ( \omega_{b}-1 \right )\left ( \eta-\eta_{in} \right ) \right ]} \right \}+\left ( \frac{1-\varsigma^2}{2\varsigma^2} \right )\left ( \eta-\eta_{in} \right ).
\end{eqnarray}

\subsection{Emission spectra }

The electron's trajectory $\boldsymbol{r}$ and the normalized velocity $\boldsymbol{\beta}$ can be used to get the radiation spectrum by the accelerated electrons. The radiation energy emitted per unit solid angle $d\Omega$ and per unit frequency interval $d\omega$ is given by (see \cite{Jakson})
\begin{equation}
\label{eq12}\frac{d^2 I }{d\Omega d\omega }=\frac{e^2\omega^2}{4\pi ^2c}\left|\textit{\textbf{n}}\times \left [ \textit{\textbf{n}}\times\textit{\textbf{F}}(\omega) \right ]\right|.
\end{equation}
From Eq.(\ref{eq12}), we will obtain emission spectrum in unit of $\frac{e^2\omega^2}{4\pi ^2c}$ in the direction of the unit vector $\boldsymbol{n}$ (do not confuse with the magnetic resonance parameter $n$, please) per unit solid angle $d\Omega$ and per unit frequency $d\omega$ as
\begin{equation}
\label{eq13}\textit{\textbf{F}}(\omega)=\int_{-\infty }^{+\infty}dt\boldsymbol{\beta}\exp{\left \{ i\omega\left [ t-\boldsymbol{n}\cdot\boldsymbol{r}(t)/c \right ] \right \}},
\end{equation}
where the dimensionless radiation frequency and vector is denoted still as $\omega=\omega/\omega_{0}$ and $\textit{\textbf{F}}(\omega)=\omega_{0}\textit{\textbf{F}}(\omega)$.
Now the dimensionless vector is
\begin{equation}
\label{eq14}\textit{\textbf{F}}(\omega)=\frac{1}{\varsigma}\int_{-\infty }^{+\infty}d\eta\boldsymbol{p}(\eta)\exp{\left \{ i\omega\left [ \eta-\boldsymbol{n}\cdot\boldsymbol{r}(\eta)+z(\eta)\right ] \right \}}.
\end{equation}

Under the electron's periodic motion, its momentum $p(\eta)$ has period $T$ as $\boldsymbol{p}(\eta+T)=\boldsymbol{p}(\eta)$ but the displacement is $\boldsymbol{r}(\eta+T)=\boldsymbol{r}(\eta)+\boldsymbol{r}_{0}$,
where $T=\frac{2\pi}{\omega_{b}-1}=2\pi n$ is the period of the electron momenta and $\boldsymbol{r}_{0}=(0,0,T\left [\left ( \frac{na}{\varsigma} \right )^2+\frac{1}{2}\left ( \frac{1}{\varsigma^2}-1 \right )  \right ])$ is the drift displacement vector of the electron during one period.
Note that the dimensionless vector $\textit{\textbf{F}}(\omega)$ can be expanded as an infinite series of delta function at the harmonics of the fundamental frequency of the emission spectrum \cite{POP-9-4325} as
\begin{equation}
\label{eq15}\textit{\textbf{F}}(\omega)=\frac{1}{\varsigma}\sum_{m=-\infty}^{+\infty}\textit{\textbf{F}}_{m}\delta\left (\omega-m\omega_{1} \right ),
\end{equation}
where the fundamental frequency in dimensionless form is given by
\begin{equation}
\label{eq16}
\omega_{1}=\frac{2\pi}{T-\boldsymbol{n}\cdot\boldsymbol{r}_{0}+z_{0}}
\end{equation}
and
the $m^{\rm{th}}$ amplitude is
\begin{equation}
\label{eq17}\textit{\textbf{F}}_{m}=\omega_{1}\int_{\eta_{in}}^{\eta_{in}+T}d\eta\boldsymbol{p}(\eta)\exp{\left \{ im2\pi h(\eta) \right \}},
\end{equation}
with
$$
h(\eta)=\frac{\eta-\boldsymbol{n}\cdot\boldsymbol{r}(\eta)+z(\eta)}{T-\boldsymbol{n}\cdot\boldsymbol{r}_{0}+z_{0}}.
$$

Now we can give some simple discussions on the relation between the resonant parameter $n$ and the fundament frequency $\omega_1$ in the case of backscatter when $\boldsymbol{n}=(0,0,-1)$.
By substituting $T=2\pi n$ and $\boldsymbol{r}_{0}$ into Eq.(\ref{eq16}), the backscatter fundamental frequency $\omega_{1}=\frac{\varsigma^2}{n+2n^3a^2}$ can lead to two limiting cases of $\omega_{1} \sim \frac{1}{n^3a^2}$ when $(na)^2\gg1$ and $\omega_{1}\sim \frac{1}{n}$ when $(na)^2\ll1$. Obviously under the given laser intensity, the larger $n$ is, i.e., strong resonance parameter, the easier the condition of $(na)^2\gg1$ satisfies. Thus when the laser intensity is given the backscattered spectrum can be adjusted to the high frequency regime by choosing the appropriate magnetic resonance parameter $n$.

Now let us focus our attention onto the $\textit{\textbf{F}}_{m}$ and the corresponding emission intensity $I_m$, we get the radiation spectrum as
\begin{equation}
\label{eq18}\frac{d^2 I_m}{d\Omega d\omega}=\frac{e^2}{4\pi^2c}\frac{1}{\varsigma^2}(m\omega_{1})^2(|\textit{\textbf{F}}_{mx}|^2+|\textit{\textbf{F}}_{my}|^2),\\
\end{equation}
where
\begin{equation}
\label{eq19}\textit{\textbf{F}}_{mx,my}=\omega_{1}\int_{\eta_{in}}^{\eta_{in}+T}d\eta\boldsymbol{p}_{x,y}(\eta)\exp{\left \{ im\omega_{1}\left ( \eta+2z \right ) \right \}}.
\end{equation}
In order to specifically analyze the typical features of the Thomson backscattered spectrum, we employed the integral expression of Integer order Bessel function $J_n (x)=\frac{1}{2\pi}\int_{0}^{2\pi}\exp{\left [  i\left ( n\xi -x\sin{\xi} \right ) \right ]}d\xi$ into Eq.(\ref{eq18}) and Eq.(\ref{eq19}) and for simplicity the assumption that the resonance parameter $n$ is an integer number is also made. Then the analytic expression of radiation spectrum is given now by
\begin{equation}
\label{eq20}\frac{d^2 I_{m}}{d\Omega d\omega}=\frac{e^2}{4\pi^2c}\frac{1}{\varsigma^2}\left ( m\omega_{1} \right )^2\left (\pi m\omega_{1}n^2a \right )^{2}\left ( A_{\left ( m,n \right )}^2+B_{\left ( m,n \right )}^2 \right )
\end{equation}
with
\begin{equation}
\label{eq21}
A_{\left ( m,n \right )}=J_{m-n}(z_{m})-J_{m+n}(z_{m})+J_{m+n+1}(z_{m})-J_{m-n-1}(z_{m}),
\end{equation}
and
\begin{equation}
\label{eq22}
B_{\left ( m,n \right )}=J_{m+n}(z_{m})+J_{m-n}(z_{m})-J_{m+n+1}(z_{m})-J_{m-n-1}(z_{m}),
\end{equation}
where we define $z_{m}=m\omega_{1}\frac{2n^3a^2}{\varsigma^2}=\frac{2mn^3a^2}{n+2n^3a^2}$ in order to simplify the formula of radiation spectrum.

While we do not intend to study the forward scattering in this paper it is worthy to say a few words about it. Obviously from  Eq.(\ref{eq16}) the fundamental frequency of forward scattering spectrum is $\omega_{1}=\frac{2\pi}{T}=1/n$. By contrast to the back direction, the forward scattered fundamental frequency is independent of laser intensity $a$. So the red shift appear in the backscattered spectrum with the increase of the laser intensity is absent in the forward scattered spectrum. On the other hand, the frequency of spectrum decrease in high degree of electron resonance ($n\gg1$) occurs in both of backward as well forward scattering.

\section{numerical results and analysis}\label{numeric}

\subsection{Thomson backward scattering spectra and its scale invariance}

\begin{figure}[htbp]\suppressfloats
\includegraphics[width=15cm]{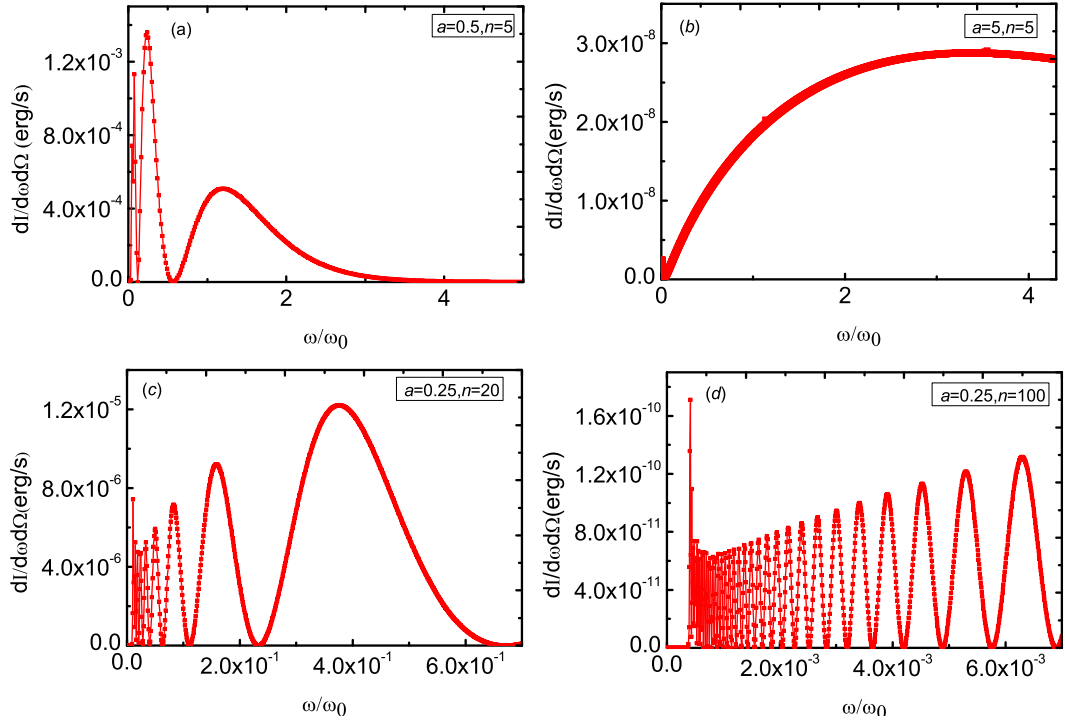}
\caption{\label{Fig1}(color online) The backward spectrum of Thomson scattering with the radiation spectrum normalized by $e^2\omega^2/4\pi^2c$. The wave length $\lambda=1\mu m$ and the initial axial momentum $p_{z0}=1$. (a) $a=0.5$, $n=5$; (b) $a=5$, $n=5$; (c) $a=0.25$; $n=20$; and (d) $a=0.25$, $n=100$.}
\end{figure}

The Thomson backward scattering spectra of the moving electron in laser and magnetic fields are presented in Fig. \ref{Fig1} for different laser field strength and resonance parameter as (a) $a=0.5$, $n=5$; (b) $a=5$, $n=5$; (c) $a=0.25$, $n=20$; and (d) $a=0.25$, $n=100$. From (a)  and (b) of Fig. \ref{Fig1}, one see that when the laser intensity $a$ increases from $0.5$ to $5$, the emission intensity of low frequency spectra decreases sharply with the increase of the laser intensity. However, the high frequency part of emission will increase while the radiation strength has decreased and the saturation effect of the backscattered radiation in Ref.\cite{PRA-60-3276} is also appeared in Fig. \ref{Fig1}(b). On the other hand, given the laser intensity $a=0.25$, when the resonant parameter $n$ varies from $20$ to $100$, an apparent oscillatory feature of the spectrum appears in (c) and (d) of Fig. \ref{Fig1} that indicates the subsequent re-emission and -absorption of the electron radiation in the fields. This type of the spectra presents the enhanced resonant radiation of the electron which exchange the energy with the laser field. The oscillatory behaviors can be as an intensification as a situation of high degree of resonant emission. In addition, the intensity of the radiation also decreases when the resonance parameter is larger, i.e. the electron is closer to the resonance condition. Obviously the high frequency spectrum dominates the radiation, especially in strong resonance situation. In the all observations of Fig. \ref{Fig1}, one can conclude that the intensity of emission will be reduced when the laser intensity $a$ or the magnetic resonance parameter $n$ increases. This phenomenon should not be surprising. Because when $(na)^2 \gg 1$ the normalized energy of the electron $\gamma \sim (na)^2\gg a$,the peak emission will appear in forward scattered direction, see Ref.\cite{PRA-93-022112}.

\begin{figure}[htbp]\suppressfloats
\includegraphics[width=8cm]{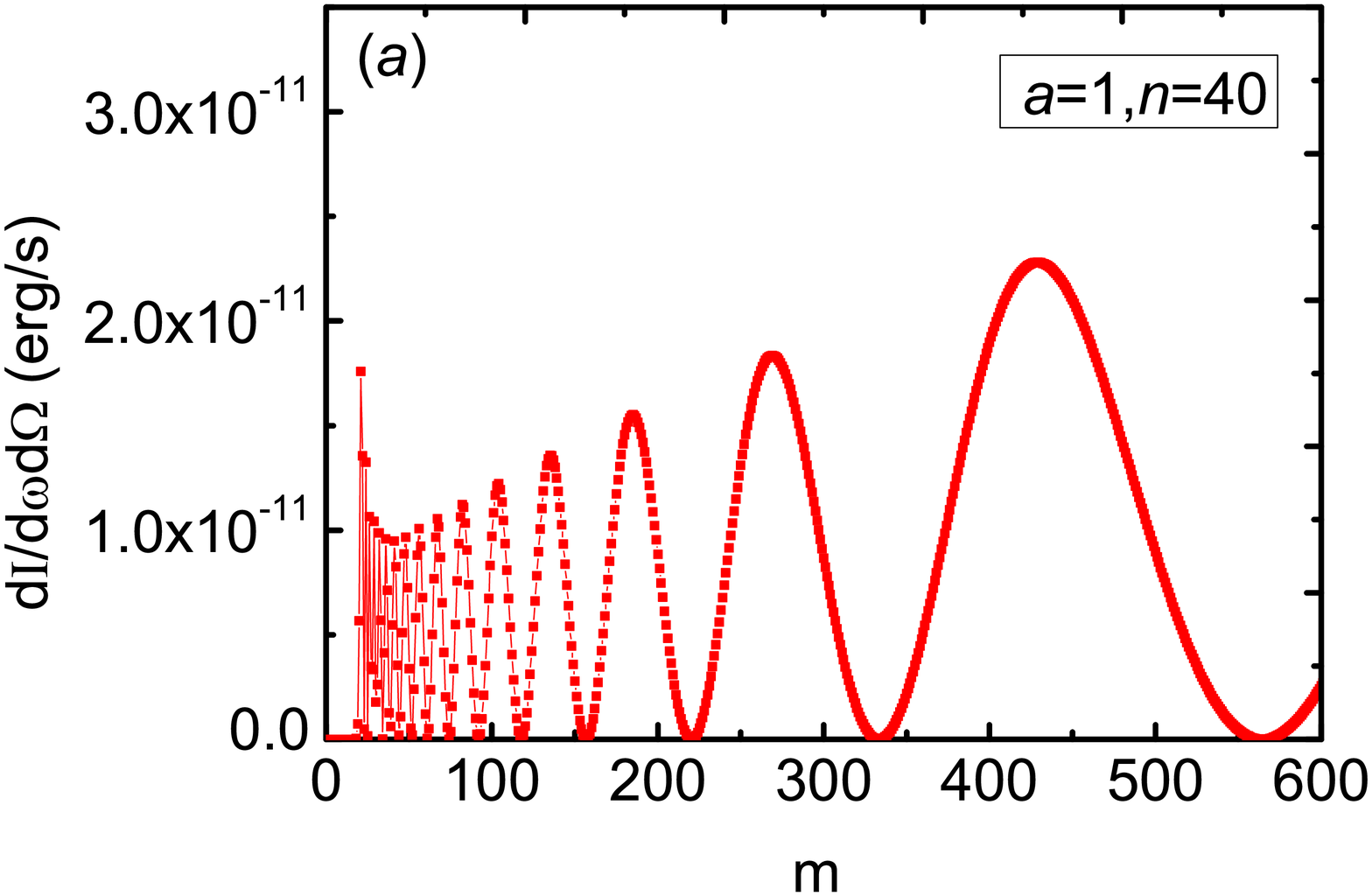}
\includegraphics[width=8cm]{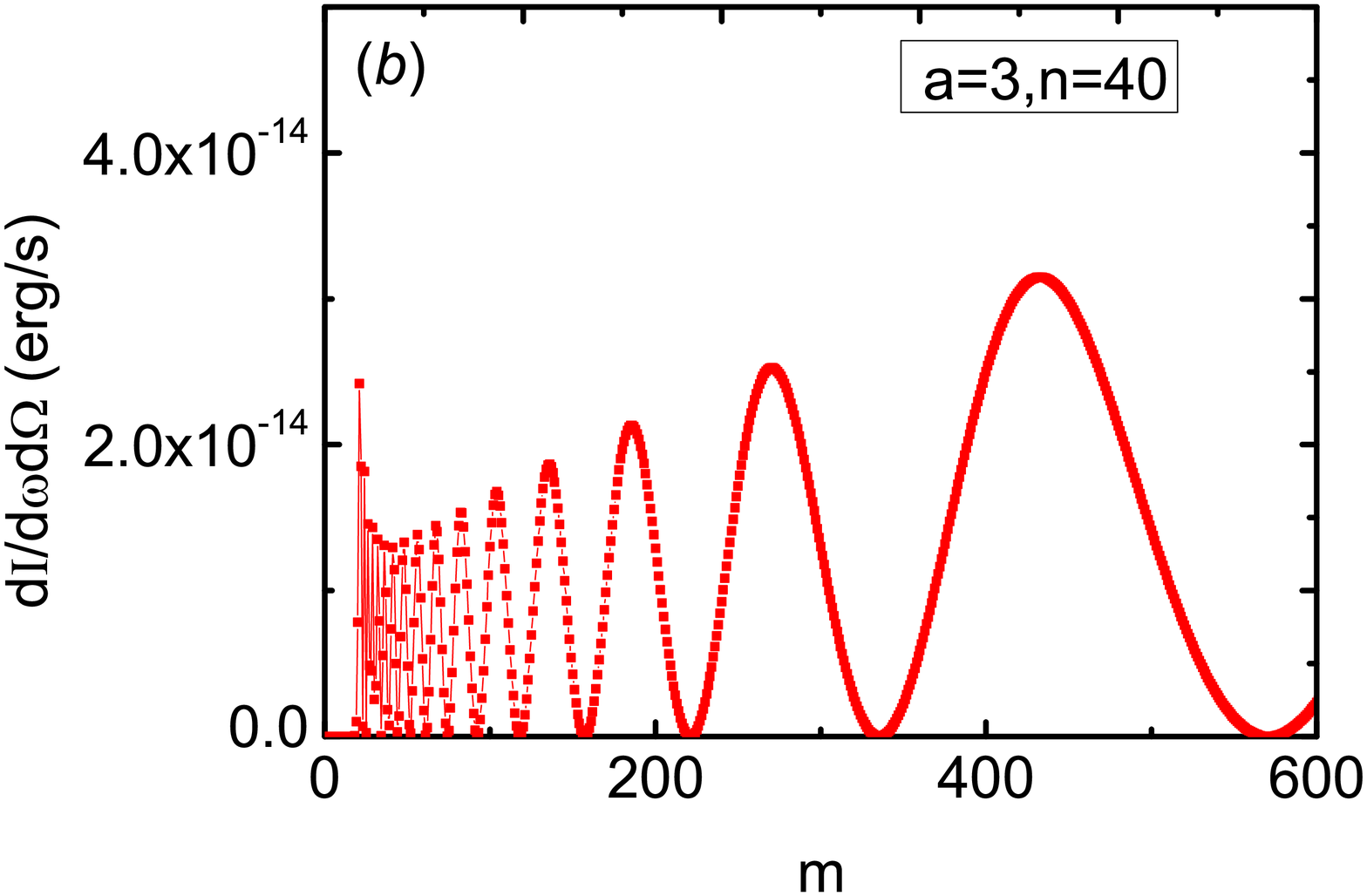}
\includegraphics[width=8cm]{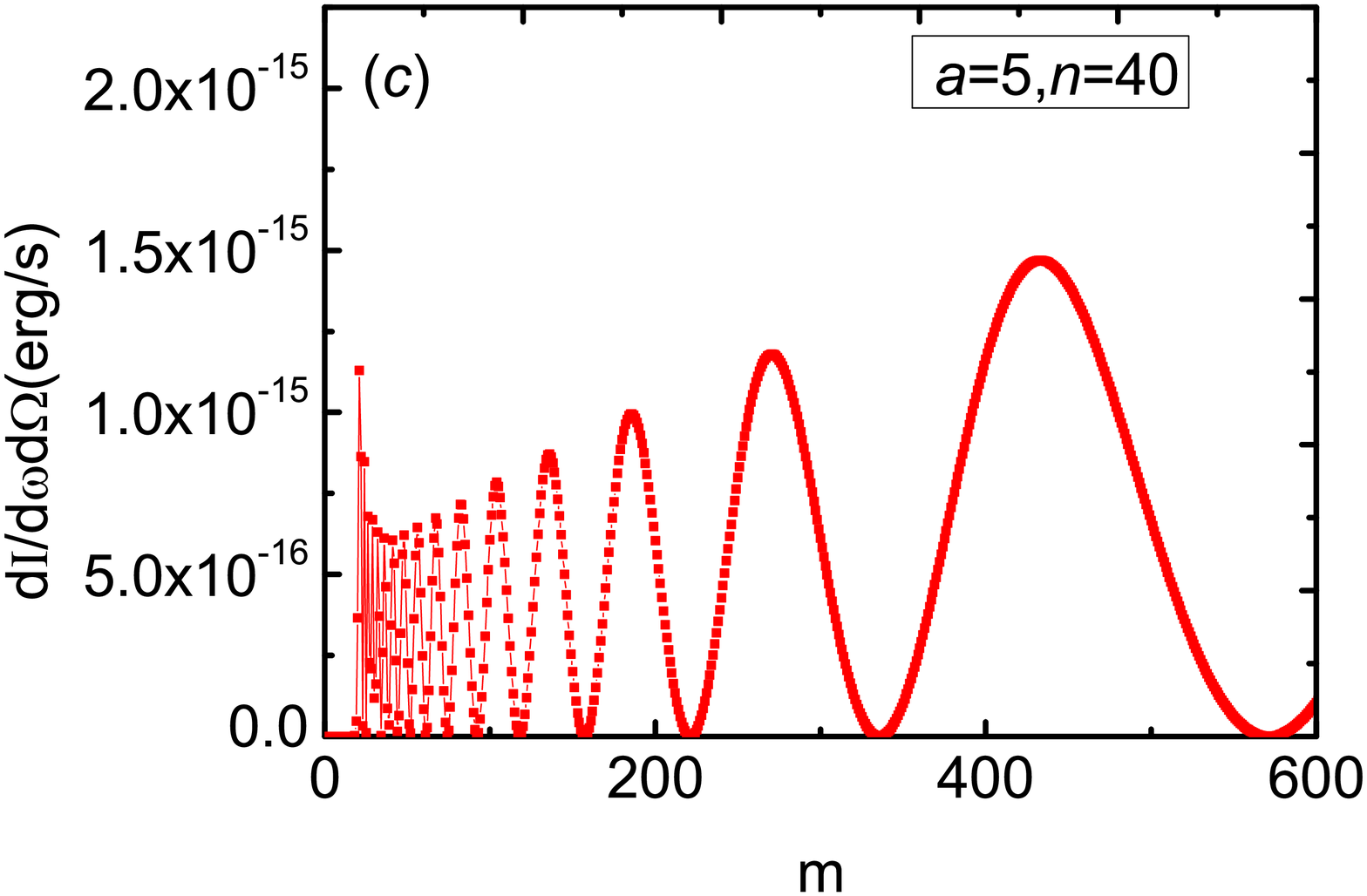}
\caption{\label{Fig2}(color online) The backscattered spectrum for harmonic orders of $\omega=m \omega_{1}$. The initial axial momentum $p_{z0}=0$. the magnetic resonance parameter $n=40$ is fixed and the laser intensities are $a=1$, $3$, and $5$ from (a) to (c), respectively.}
\end{figure}

Another interesting phenomenon we found is the scale invariance of the spectrum for either the laser intensity $a$ in high degree resonance regime where magnetic resonance parameter $n\gg1$ or the initial electron axial momentum $p_{z0}$. In the former case some of the backscattered spectrum is calculated numerically for integral orders of harmonic from $0$ to $600$, which are shown in Fig. \ref{Fig2}, where the magnetic resonance parameter $n=40$ is chosen and fixed and the laser intensities are $a=1$, $3$, and $5$ from (a) to (c), respectively. Obviously the shape of backward scattering spectrum remains unchanged almost except that the intensity of the spectrum will decrease as laser intensity $a$ increases.

\begin{figure}[htbp]\suppressfloats
\includegraphics[width=15cm]{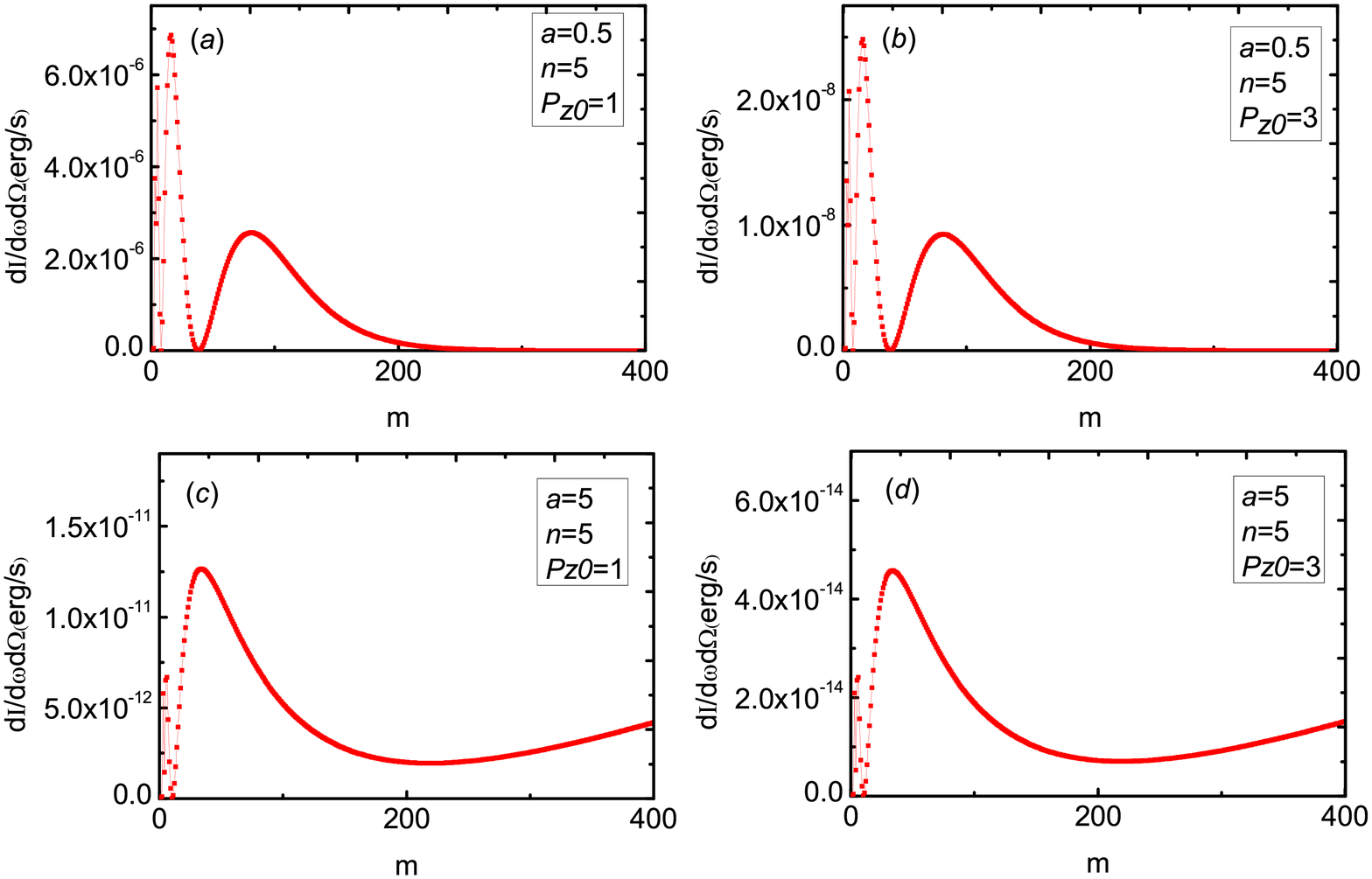}
\caption{\label{Fig3}(color online) The backward spectra of $m^{\mathrm{th}}$ harmonic orders with different parameters. (a)  $a=0.5$, $n=5$, $p_{z0}=1$; (b) $a=0.5$, $n=5$, $p_{z0}=3$; (c) $a=5$, $n=5$, $p_{z0}=1$; and (d) $a=5$, $n=5$, $p_{z0}=3$.}
\end{figure}

We also research the relationship between the spectrum of the radiation and the axial initial momentum of the electron. The spectrum of  integer orders of harmonic is plotted in Fig. \ref{Fig3} for four sets of different parameters as (a)  $a=0.5$, $n=5$, $p_{z0}=1$; (b) $a=0.5$, $n=5$, $p_{z0}=3$; (c) $a=5$, $n=5$, $p_{z0}=1$; and (d) $a=5$, $n=5$, $p_{z0}=3$. It is found that first the same effect of saturation also appear in high frequency regime of spectra while the low frequency strength of spectrum falls down sharply. Second all radiation intensities decrease rapidly with the increase of the laser intensity. Third, it may be the significant result, the shape of Thomson scattering spectrum is almost independent of its axial initial momentum, which can be evident by make a comparison between Figs.\ref{Fig3} (a) and (b) as well as Figs.\ref{Fig3} (c) and (d). By the way from Figs.\ref{Fig3} (a) and (c) or/and Figs.\ref{Fig3} (b) and (d) one can see that the scale invariance about the laser intensity disappear almost when other parameters are fixed. The reason is that the $n=5$ does not lies in the high degree resonance regime. Moreover it seems lack of scale invariance about $n$ when $a$ and $p_{z0}$ are fixed because the $n$ appear not only in the argument variables but also the orders of Bessel functions, see Eqs.(\ref{eq21}-\ref{eq22}).

From results aforementioned it is found that the scale invariance of the spectrum exists in the Thomson backscatter spectra. This scale invariance implies also that the intensity of Thomson backscattered spectrum is inversely proportional to some powers of the laser intensity $a$ and the initial axial momentum $p_{z0}$ for given special $m^{\rm{th}}$ harmonic. This intrigues directly the following study about the scaling law.

\begin{figure}[htbp]\suppressfloats
\includegraphics[width=10cm]{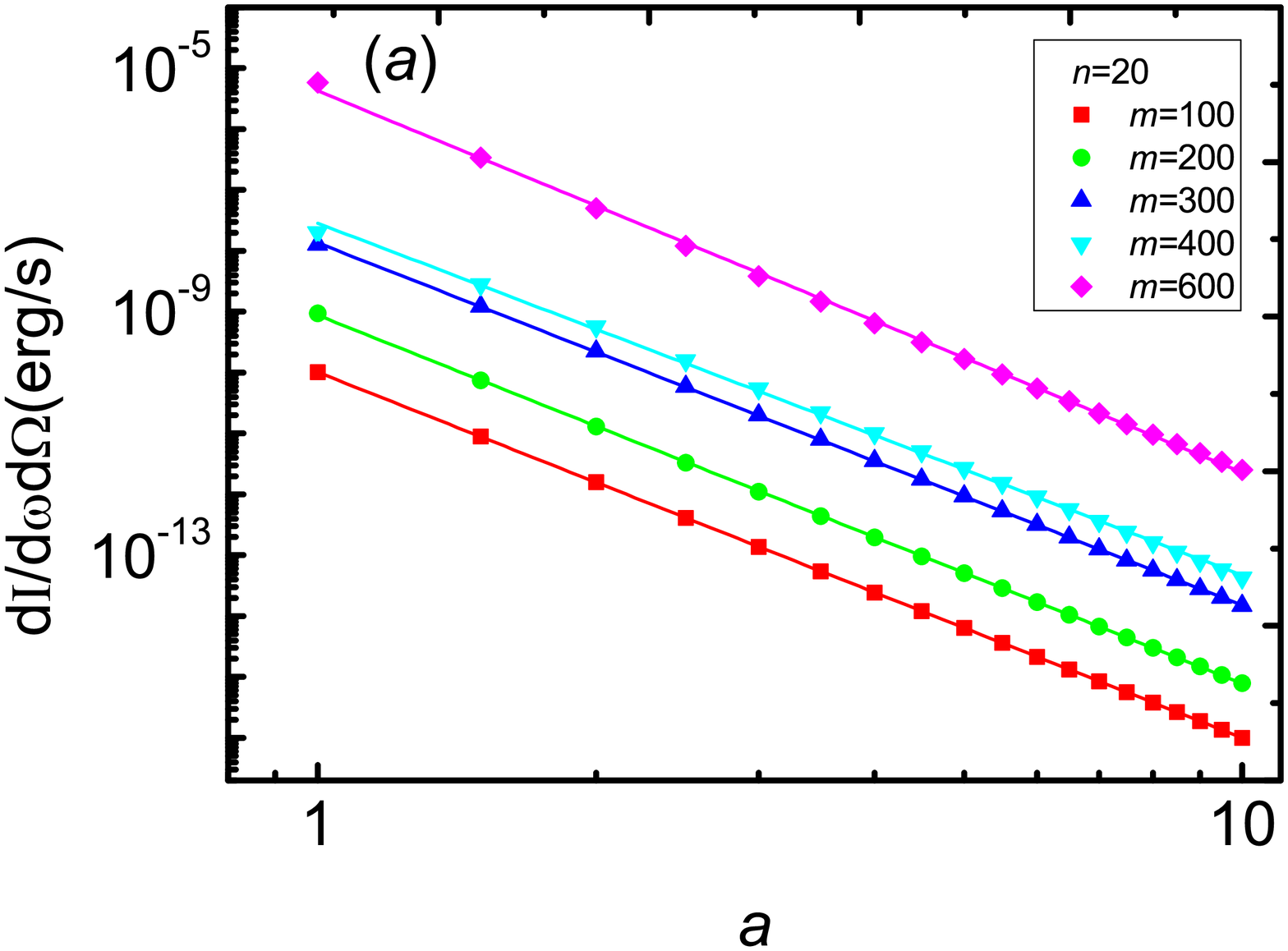} \\
\includegraphics[width=10cm]{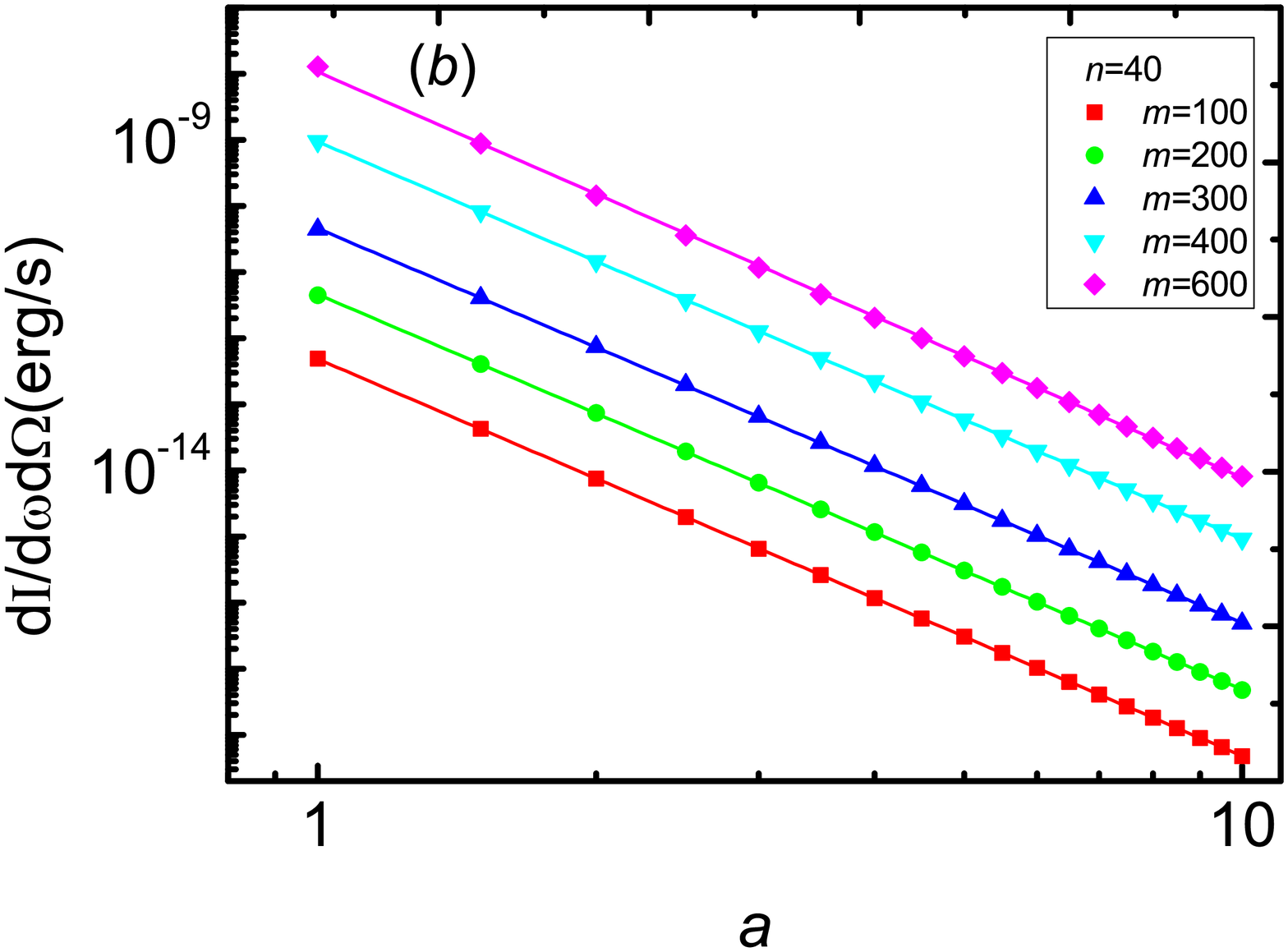} \\
\includegraphics[width=10cm]{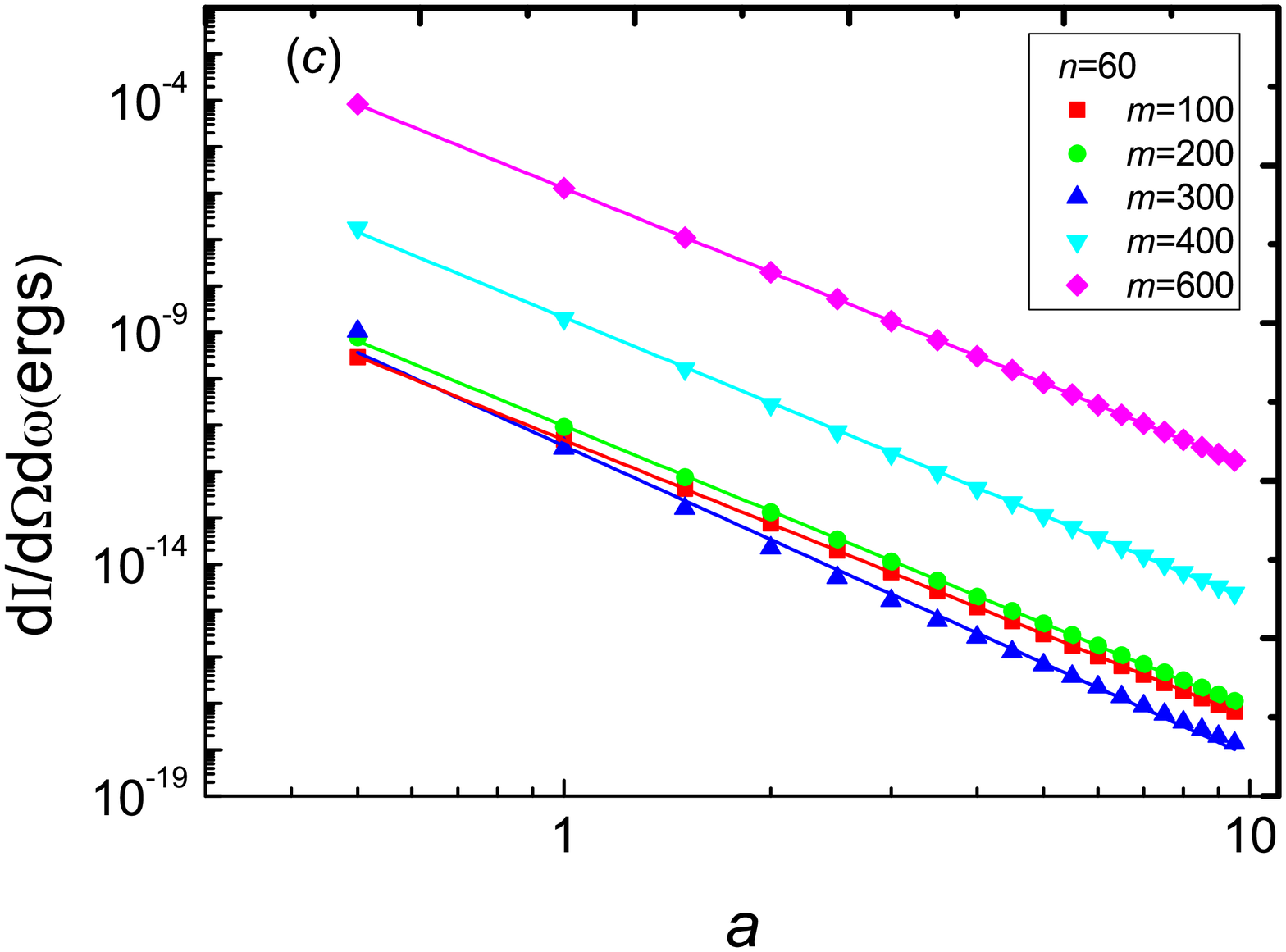}
\caption{\label{Fig4}(color online) The scaling law of the backscattered spectrum with the intensity of the laser field $a$ for different harmonic orders of $m=100$, $200$, $300$, $400$, and $600$. The intensity of the emission multiplied by $10,10^2,10^3,10^4,10^6$ successively. From top to bottom the resonance parameters are $n=20$, $40$, and $60$.}
\end{figure}

\begin{figure}[htbp]\suppressfloats
\includegraphics[width=15cm]{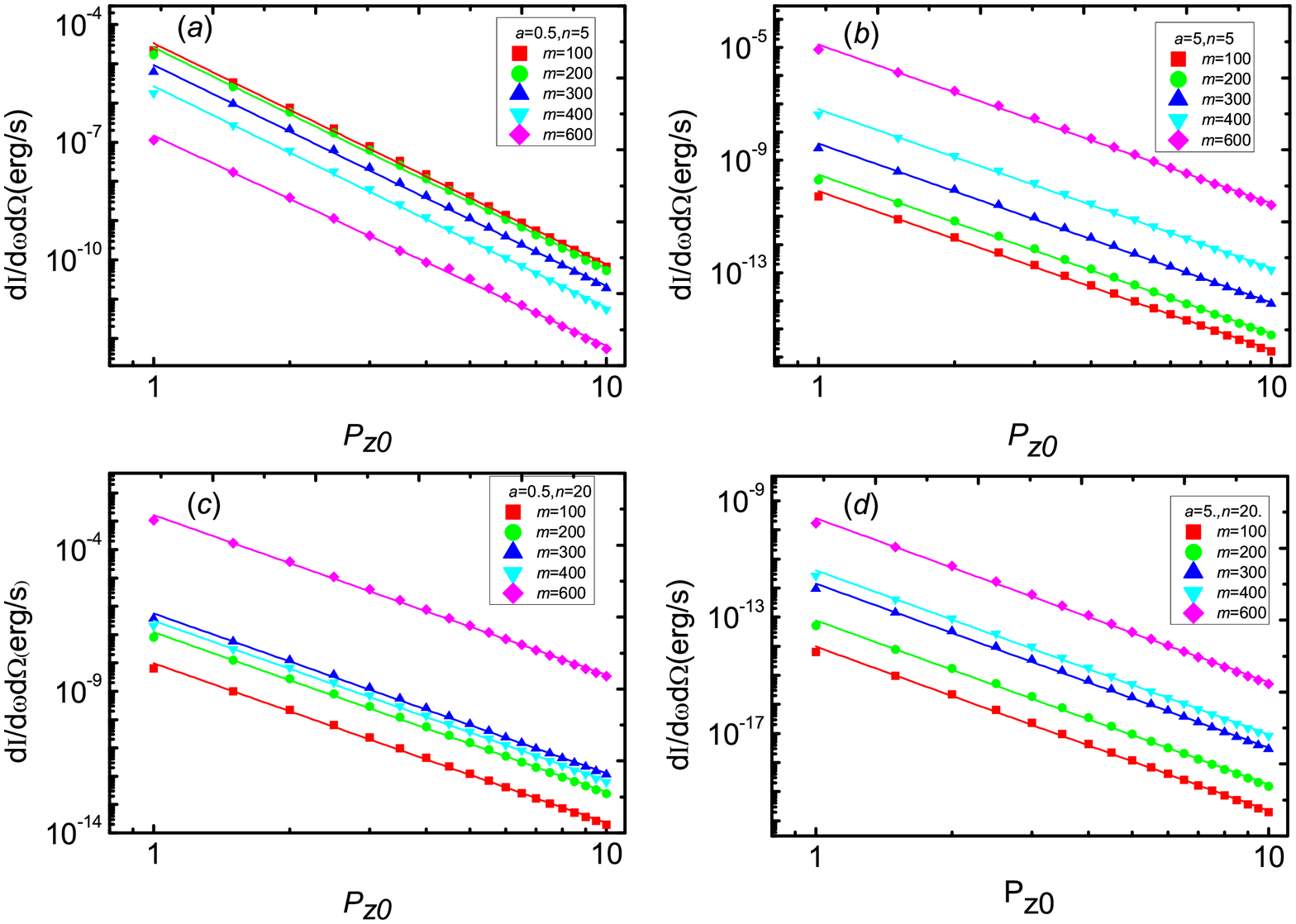}
\caption{\label{Fig5}(color online) The scaling law of the backscattered spectrum with the initial axial momentum $p_{z0}$ for different harmonic orders of $m=100$, $200$, $300$, $400$, and $600$. The intensity of the emission multiplied by $10,10^2,10^3,10^4,10^6$ successively. (a) $a=0.5$, $n=5$; (b) $a=5.0$, $n=5$; (c) $a=0.5$, $n=20$; and (d) $a=5$, $n=20$.}
\end{figure}

\subsection{The scaling law of Thomson backward scattering spectra}

The scaling law of the backward radiation with respect to the intensity of the laser field is observed when it varies from $a=1$ to $a=10$ in high degree cases of magnetic resonance. In Fig. \ref{Fig4}, we choose harmonic orders of $m=100$, $200$, $300$, $400$ and $600$ that reveals its feature apparently. The magnetic resonance parameters are chosen as (a) $n=20$; (b) $n=40$ and (c) $n=60$, respectively, reminded by the scale invariance of the spectrum exists in high resonance region from previous discussion. On one hand it is known that the intensity of the spectra will decrease with the increase of resonance parameter which is in accordance with Fig. \ref{Fig3}. On the other hand the intensity of the spectrum is proportional to the laser intensity $a^{-6}$ in high resonance regime where $n\gg1$. This scaling law concerning $a$ is also in accordance with Fig. \ref{Fig2}. When the laser intensity $a$ changes from $1$ to $10$, the scaling law always holds when the high magnetic resonance parameter is satisfied.

Similarly, Fig. \ref{Fig3}, which displays the scale invariance of the initial axial momentum, reminds us that the existence of the scaling law of $p_{z0}$.  In Fig. \ref{Fig5} we show the results of the backward radiation for harmonic orders of $m=100$, $200$, $300$, $400$ and $600$, where $p_{z0}$ is changed from $1$ to $10$ but the other parameters are chosen as (a) $a=0.5$, $n=5$; (b) $a=5.0$, $n=5$; (c) $a=0.5$, $n=20$ and (d) $a=5$, $n=20$. Note that in order to distinguish some of the very approaching curves the intensity of radiation is multiplied by $10^{m/100}$ for each $m$. By comparing with the scaling law of $a$, the pronounced difference is that the scaling law of initial axial momentum of electron can be maintained even if the magnetic parameter is not high ($n \gg 1$) as $n=5$, see Fig. \ref{Fig5}(a) and(b). No matter how the laser intensity $a$ or magnetic resonance parameter $n$ changes, the scaling law of the spectra about the electron initial axial momentum remains always good. This also indicates that the magnetic resonance parameter and the laser intensity are independent factors to create the scaling law of the initial axial momentum.

The theoretical interpretation of the observed scaling law above can be derived from Eqs.(\ref{eq20}-\ref{eq22}) in terms of the analytic formula of the Thomson backscattered spectrum associated to the Bessel functions. When the magnetic resonance parameter $n\gg1$, then the parameter $z=m\omega_{1}\frac{2n^3a^2}{\varsigma^2}=\frac{2mn^3a^2}{n+2n^3a^2}$ will be simplified to $z=m$, so that the value of Bessel function $J_{n}(z)$ is independent of $a$ and $n$, which will present the invariance of the spectrum on account of harmonic orders in Fig. \ref{Fig2} and Fig. \ref{Fig3}. It leads that the shape of the spectra is invariable when the magnetic resonance parameter $n\gg1$ which corresponds to the high resonance region.The scaling law of the magnetic resonance parameter $n$ is undiscovered,because $n$ can affect the order of Bessel function from the analytic results.

On the other hand, the intensity of the backscattered spectrum is proportional to $\omega_{1}^4 n^4 a^2/\varsigma^2$. The fundamental frequency $\omega_{1}=\varsigma^2/(n+2n^3 a^2)$ can be substituted into the proportional relationship. In the condition of $p_{z0}\gg1$ we have $\sqrt{1+p_{z0}^2} \approx p_{z0} +1/ 2p_{z0}$ so that $\varsigma \approx 1/p_{z0}$, therefore, we derive the scaling relationship of the radiation as $d^2 I_{m}/ d\Omega d\omega \sim (a p_{z0})^{-6}$. This negative sixth power law of the laser intensity $a$ and the initial axial momentum$p_{z0}$ explains analytically for the scaling laws observed in Fig. \ref{Fig4} and Fig. \ref{Fig5} numerically. It is worthy to point out that this power law relationship reveals the effects of the laser intensity and the magnetic resonance acceleration electron on the emission of the radiation, which has not been unveiled in previous works.

\section{conclusions and discussions}

In this study the Thomson backscattered spectrum of the electron has been explored in the framework of laser-magnetic resonance acceleration. When the electron is close to the resonance motion regime, i.e., the magnetic resonance parameter $n\gg1$, the intensive re-emission and -absorption can not only lead to a series of peaks of backscattered spectra but also result in the saturation of the high frequency spectra. The backward scatter fundamental frequency under combined laser and magnetic fields can be simplified as $\omega_{1}\sim\frac{1}{n^3a^2}$ when $(na)^2\gg1$ and $\omega_{1}\sim\frac{1}{n}$ when $(na)^2\ll1$. For high resonance regime, we can adjust the degree of magnetic resonance to move the backscattered spectrum towards high frequency region with the same laser intensity.

The scale invariance of the spectrum based on harmonic orders $m$ is found, when the laser intensity $a$ and initial axial momentum $p_{z0}$ are considered as scale factors. The absence of $n$ as the scale factor is that it can not only appear in the argument of functions but also appeared as the order of Bessel functions.
The spectrum will maintain its shape with different laser intensity in high resonance regime and initial axial momentum. The only difference is the decrease of radiation as the increasing of laser intensity and axial initial momentum. We also obtain the analytic formula of the radiation emission for the sake of quantitatively explaining its feature. The scaling law of the radiation is derived as $\frac{d^2 I_{m}}{d\Omega d\omega}\sim\frac{1}{a^6 p_{z0}^6}$ in the condition of $n\gg1$ and $p_{z0}\gg1$, which means that the intensity of emission is inversely proportional to the sixth power of the laser intensity $a$ and to the initial axial momentum $p_{z0}$. This fact accords with the scale invariance of backscattered spectrum. These interesting results reveal the possibility of tunable and compact advance x-ray source. Especially when the laser field intensity is limited we can adjust the magnetic field as an alternative way to give higher frequency of harmonic radiation.

In present study some questions are not included so that the involved problem is still open. For example, the classical model we use in this paper excludes the radiation reaction effect. The different configuration of external magnetic field maybe produce the different effect to the spectrum. In addition, the forward scattering spectrum is also altered by magnetic resonance effect which is not studied in present paper. Recently the involvement of quantum effect has been considered by the total quantum description of Compton scattering in Ref. \cite{PRD-41-043014}. Obviously these impact factors beyond the scope of present study and worthy to be analyzed and studied in the future.

\begin{acknowledgments}
This work was supported by the National Natural Science Foundation of China (NSFC) under Grant Nos. 11475026, 11305010 and 11175023. The computation was carried out at the HSCC of the Beijing Normal University. The computation was carried out at the HSCC of the Beijing Normal University.
\end{acknowledgments}

\end{document}